\documentclass{article}
\usepackage{arxiv}
\usepackage[T1]{fontenc}    
\usepackage{hyperref}       
\usepackage{url}            
\usepackage{booktabs}       
\usepackage{amsfonts}       
\usepackage{nicefrac}       
\usepackage{microtype}      
\usepackage{lipsum}
\usepackage{algpseudocode}
\usepackage{amsmath}
\usepackage{float}
\usepackage{amsmath,amssymb,amsfonts}
\usepackage{multirow}
\usepackage{algorithm}
\usepackage{graphicx}
\usepackage{xcolor}
\usepackage{setspace}
\usepackage[algo2e,linesnumbered,lined,boxed,commentsnumbered, ruled]{algorithm2e}
\usepackage[style=ieee,backend=biber]{biblatex}
\graphicspath{ {./images/} }

\title{Epidemic Forecasting with a Hybrid Deep Learning Method Using CNN-LSTM With WOA-GWO Parameter Optimization: Global COVID-19 Case Study}
\author{
  Mousa Alizadeh \\
  School of Engineering, RMIT \\
  Melbourne, Australia \\
  \And
    Mohammad Hossein Samaei* \\
  Tarbiat Modares University \\
  Tehran, Iran \\
  \texttt{*corresponding author:}\\
  msamaei@ksu.edu \\
    \And
  Azam Seilsepour \\
  Islamic Azad University,\\
  Central Tehran Branch \\
  Tehran, Iran \\
  \And
    Alireza Monavarian \\
  Ferdowsi University of Mashhad \\
  Mashhad, Iran \\
  \And
  Mohammad TH Beheshti \\
  Tarbiat Modares University \\
  Tehran, Iran \\
}
\begin{document}
\maketitle
\begin{abstract}
Effective epidemic modeling is essential for managing public health crises, requiring robust methods to predict disease spread and optimize resource allocation. This study introduces a novel deep learning framework that advances time-series forecasting for infectious diseases, with its application to COVID-19 data as a critical case study. Our hybrid approach integrates Convolutional Neural Networks (CNNs) and Long Short-Term Memory (LSTM) models to capture spatial and temporal dynamics of disease transmission across diverse regions. The CNN extracts spatial features from raw epidemiological data, while the LSTM models temporal patterns, yielding precise and adaptable predictions. To maximize performance, we employ a hybrid optimization strategy combining the Whale Optimization Algorithm (WOA) and Gray Wolf Optimization (GWO) to fine-tune hyperparameters, such as learning rates, batch sizes, and training epochs—enhancing model efficiency and accuracy. Applied to COVID-19 case data from 24 countries across six continents, our method outperforms established benchmarks, including ARIMA and standalone LSTM models, with statistically significant gains in predictive accuracy (e.g., reduced RMSE). This framework demonstrates its potential as a versatile method for forecasting epidemic trends, offering insights for resource planning and decision-making in both historical contexts, like the COVID-19 pandemic, and future outbreaks.
\end {abstract}


\maketitle

\section{Introduction}
COVID-19 affected the lives of people all over the world and took the lives of thousands around the world. On the other hand, the social and economic consequences of COVID-19 sometimes inflicted irreparable damage on countries, and it was able to confront the world with new challenges and crises. Therefore, many researchers have tried to minimize the damage caused by this disease by providing models to predict the rate of coronary artery disease. Various approaches to forecasting have also been developed. For instance, \cite{He2020ADS} developed a discrete-time stochastic model to describe the dynamics of pandemic expansion. Prediction Models Traditional time series have been routinely studied to predict COVID-19 cases \cite{kucharski2020early, wu2020nowcasting, zhuang2020preliminary}. However, Stochastic modeling has some drawbacks such as being incapable of handling large data scales, modeling complex nonlinear characteristics, and its dependence on pre-defined features or assumptions\cite{ihme2020forecasting}. Furthermore, the traditional Autoregressive Integrated Moving Average (ARIMA) model proved to be a suitable approach for describing short-term autocorrelation in time series data, researchers have commonly used it. \cite{collou2019digitalization}  utilized a  time series method such as the ARIMA to predict the number of validated items and compared to other conventional methods such as stochastic modeling, which ARIMA proved to be more accurate in short-time forecasting. Other studies have been developed in India using traditional ARIMA modeling \cite{gupta2020trend}. In 2022, other researchers from Bangladesh used ARIMA to predict verified people, deaths, and recoveries in COVID-19 \cite{nesa2022forecasting}. \cite{ribeiro2020short} used ARIMA, cubist regression (CUBIST), random forest (RF), Ridge Regression (RIDGE), Support Vector Regression (SVR), and stacking-ensemble learning to forecast the total number of patients in Brazil in 1, 3, and 6 days.  ARIMA has had superior performance in the case of short-time forecasting. There are also some other studies devoted to developing ARIMA approaches for forecasting the Covid epidemy \cite{roy2021spatial,kufel2020arima,singh2020development,moftakhar2020exponentially, liu2020predicting, duan2020arima, arunkumar2021forecasting}.\\

While ARIMA may provide a good fit for the initial spread of COVID-19, its effectiveness may diminish as the pandemic progresses and new interventions are implemented \cite{hernandez2020forecasting}. On the other hand, another approach, namely logistic function, may allow for a more flexible and dynamic approach to modeling the pandemic's impact on a particular population. There are studies on developing a logistic function to predict the spread of COVID-19 \cite{roosa2020real}. The logistic function method assumes that the spread of COVID-19 can be modeled using a sigmoidal curve, which allows for the prediction of an eventual plateau or decline in the number of cases, which are discussed completely in \cite{chen2020reconstructing,li2020trend,qeadan2020naive} which proved to be an appropriate approach when the number of cases is decreasing.\\

In \cite{wang2020prediction} a more advanced algorithm named Logistic Patient Information Based Algorithm (LPIBA) to estimate the number of COVID-19-related deaths in China. Their results show that the overall mortality rate in Hubei and Wuhan is 13 percent and between 0.75 percent and 3 percent in the rest of China. Furthermore, in \cite{fathollahi2022sustainable}, the LPIBA was developed to address forecasting the infected cases. The LPIBA proved to have superior performance over ARIMA, having features such as Incorporation of patient information, flexibility in terms of data type, and ability to handle binary outcomes. However, it is worth mentioning the LPIBA has limitations when detailed data is not available or it is not feasible to collect some features. 
\cite{mohan2022predicting} used a hybrid ARIMA and Prophet Model to predict daily confirmed and cumulative confirmed cases in India. The Prophet model is open-source developed by Facebook Data and proved to have appropriate performance with time-series types of data having seasonal effects \cite{taylor2018forecasting,taylor2017prophet}. It proved to be a robust algorithm in the case when there are missing data \cite{ndiaye2020analysis}. In another work, Hosseini et al. proposed a Bayesian predictive method to predict the West Nile virus cases\cite{hosseini2024parsimonious}, however, these methods also need careful parameter selection and prior distribution assumptions.
\\

Another important and popular approach to epidemic modeling involves treating disease spread as stochastic processes over networks \cite{pastor2015epidemic}. Historically, these methods were constrained to small populations, typically on the order of a few thousand individuals. However, the recently introduced FastGEMF module has enabled simulations encompassing millions of nodes over multi or single-layer complex networks \cite{10876117}. While network-based disease spread models provide precise insights into disease propagation, their stochastic nature and the high degree of freedom (DOF) in selecting network parameters or disease models—such as SIS, SIR, and others—often necessitate extensive simulations. This approach is particularly suitable when substantial information is available regarding disease dynamics or population structure, as demonstrated in studies such as \cite{das2024sir} that model spill-over using the SIR model over interconnected networks with a case study of the Ebola virus to see how the level of interconnection can lead to an epidemic in human networks. In contrast, for newly emerging diseases like COVID-19, purely data-driven methods are often more appropriate \cite{XU2022105342}.\\

To deal with more complex data, Neural Networks proved to have superior performance over Prophet Model and ARIMA\cite{bengio2013representation,schmidhuber2015deep,tygert2016mathematical}. In \cite{de2022covid}, authors used the ANN algorithm to forecast the number of cumulative cases of infection and deaths in Brazil. They also used a substantial mitigation procedure adopted (mandatory use of masks) was experimented as an input to evaluate the improvement in the results. \cite{vadyala2021prediction} developed a more complex NN, a hybrid algorithm for predicting short-term COVID-19 cases in Louisiana, USA. Their proposed algorithm combines selecting similar day features using Xgboost, K-Means, and LSTM. In \cite{arora2020prediction}, researchers in India compared the Convolutional LSTM, BILSTM, and Stacked LSTM algorithms to the MAPE test to use an optimal algorithm to predict the number of coronary arteries in India Other researchers used simple Recurrent Neural Network (RNN) algorithms, LSTM, two-way LSTM (BiLSTM), Gateway Recurrent Units (GRUs), and Variable Automatic Encoder (VAE) to predict COVID-19 in six countries \cite{zeroual2020deep}.\\

Authors in \cite{kafieh2020isfahan} used time-series data related to data from COVID-19 in Iran. Using the LSTM algorithm, they sought to include the interaction of all classes of coronary arteries, deceased coronary arteries, and the number of recovered individuals in the prediction process using this method.  \cite{abbasimehr2021prediction} used the multi-head attention-based method (ATT\_BO), CNN-based method (CNN\_BO), and LSTM-based method (LSTM\_BO) to forecast short-term and long-term COVID-19. For this reason, they perform two types of datasets. \cite{achterberg2020comparing} used a Network Inference based Prediction Algorithm (NIPA) to forecast the corona in the Chinese city of Hobby and the Netherlands. They also compared their method with LSTM and sigmoid curve methods In the same years, \cite{galasso2022random} from the USA presented a relatively non-parametric random forest model to forecast the number of COVID-19 cases in the U.S. \cite{xu2022forecasting} used CNN, LSTM, and the CNN-LSTM algorithm to predict the number of COVID-19 in Brazil, India, and Russia. They also compared the performance of these models with the previously developed deep learning models. \cite{zhou2022interpretable} used the Interpretable Temporal Attention Network (ITANet) algorithm to forecast COVID-19. By using this algorithm, they wanted to infer the importance of government interventions.\\

Consequently, neural networks and deep learning algorithms proved to be a promising approach that can forecast with high accuracy and are compatible with large data sets and complex systems which have contributed to more than 40 percent of epidemy forecasting \cite{rahimi2021review} and many pieces of research other than those mentioned previously have been devoted to deep learning algorithms \cite{chimmula2020time,ayyoubzadeh2020predicting,fong2020finding,tamang2020forecasting,horby2020effect,sujath2020machine,sholehrasa2024integrating}. Although much research has been developed during the Corona to provide the best predictive results, very few studies have provided a comprehensive analysis of the combination of different neural network algorithms to optimize results and increase prediction accuracy. In other words, to use the important and practical features of different algorithms, they can be combined together to get better results from the models. Also, the use of Metaheuristic algorithms has rarely been investigated in this area. To this end, and to optimize the results, this study, in addition to using a hybrid neural network, for the first time combines two Metaheuristic algorithms of gray wolf and whale on the proposed neural network. Finally, the proposed model is examined on the data of twenty-four countries from six continents in the world.\\

The main contribution of this article can be summarized as: 
\begin{itemize}
    \item A comprehensive study of the number of infected, cured, and deaths of COVID-19 in 24 countries around the world, for the first time, has been presented.
    \item Presenting a   novel hybrid deep learning-based method..
    \item The hybrid meta-heuristic algorithms have been utilized for hyperparameters tuning.
    \item The proposed method of this article has the advantage of being used with different time-series type data (the generality of the proposed methods)
\end{itemize}

\section{Forecasting Method}\label{secBasicOfResearch}
This section briefly explains the proposed method's building blocks, including CNN, LSTM, and GWO-WOA. The proposed method combines CNN and LSTM for prediction and employs the GWO-WOA for hyperparameter tuning.

\subsection{Convolutional Neural Network}\label{secCNN}
\begin{figure}
	\centering
	\includegraphics [width=5.7 in] {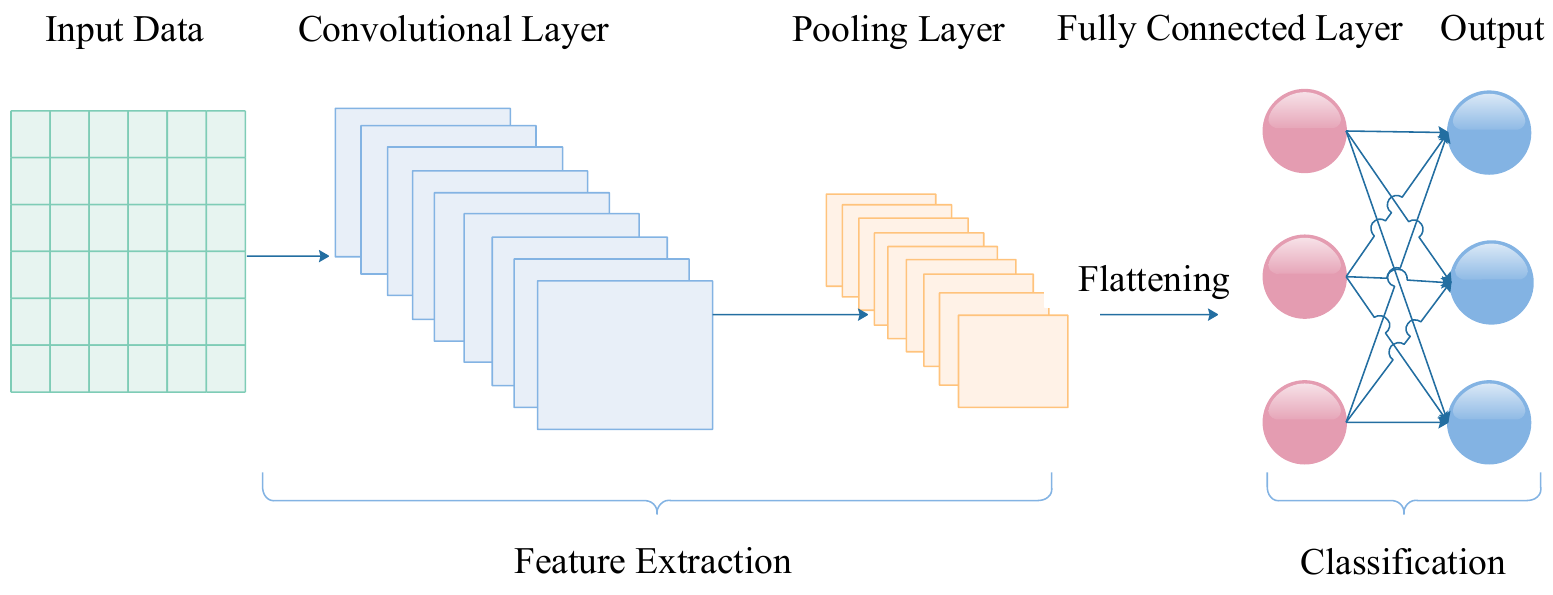}
	\caption{The CNN Structure}
	\label{1}
\end{figure}

During the 1960s, Hubel and his colleagues conducted biological research that demonstrated how visual information is transmitted from the retina to the brain through multiple levels of receptive field excitation. This discovery eventually led to the development of Convolutional Neural Networks (CNNs). A CNN is a type of feed-forward neural network composed of various layers such as input, convolution, pooling, fully connected, and output layers. The input layer processes data through feature transformation and extraction using convolution and pooling layers, which integrate local information. The fully connected layers then map this information to the output signals generated by the output layer after receiving specific properties  \cite{lecun2015deep}. The structure of a CNN is depicted in Figure 1, and its calculation formula is presented in Equation 1, where N, W, and F represent the output size, input size, and size of the kernel, respectively. Additionally, P and S denote padding size and step size, respectively.
\begin{equation}
N=\frac{W-F-2P}{S+1}
\end{equation}
The main function of the convolution layer is to extract features from the input data. Typically, the convolution layer of a CNN is made up of three main components: the convolution kernel, convolutional layer parameters, and an activation function. This layer is widely regarded as the most essential and standout component within a CNN architecture. The convolution layer in a CNN leverages convolution kernels to extract features from input variables. The size of the kernel employed has usually a smaller value compared to that of the input matrix. Unlike general matrix operations, the computation within convolutional layers utilizes The feature map output generated through convergence operations, wherein each element within the feature map is computed using Equation 2.  \cite{wang2017deterministic}. This equation denotes the output value at row $i$ column $j$ of the feature map (referred to as $x_{i,j}^{out}$)), in conjunction with the corresponding value at row $i$ column $j$  of the input matrix (denoted as $x_{i+m,j+n}^{in}$) and the weight (represented as  $w_{m,n}$ within row $m$  and column $n$ of the convolution kernel. Additionally, the bias (represented as b) associated with the convolution kernel is taken into consideration, along with the utilization of the chosen activation function (denoted as $f_{cov} (0)$ ).

\begin{equation}
x_{i,j}^{out}=f_{cov}(\sum_{m=0}^k \sum_{n=0}^k w_{m,n}x_{i+m,i+n}^{in} + b)
\end{equation}
Generally, the input matrix within the convolutional layer of Convolutional Neural Networks (CNNs) leverages multiple kernels. Each individual kernel operates by extracting distinctive features from the input matrix, subsequently generating a corresponding feature map. Subsequently, the pooling layer assumes the task of reducing the dimensions of the preceding feature map, effectively decreasing both its length and width. This down-sampling operation not only enhances computational efficiency but also contributes to overall performance improvement. The convolutional layer's output, comprising feature vectors, can be further diminished in dimensionality through the application of the pooling layer. Remarkably, this reduction process simultaneously results in enhanced outcomes. This is attributed to CNNs' inherent capability to proficiently extract features from grid-like data structures. As a result, a technique involving the expansion of m variables of arbitrary nature into n stations has been employed to obtain a matrix consisting of m rows and n columns. Thus, one can regard the CNN, as a whole, as encompassing a fully connected layer that serves as a classifier. Positioned at the terminal section of the network, this fully connected layer conducts regression classification on the features that have been extracted. Consequently, the CNN can be logically divided into two distinct parts: the initial segment involves feature extraction, encompassing operations such as convolution, activation functions, and pooling. On the other hand, the latter part of the CNN architecture focuses on classification and recognition, primarily facilitated by the fully connected layer  \cite{yao2021optimal}. For a visual representation of the CNN model's structural layout,  refer to  Figure \ref{1}. 

\subsection{Long Short-Term Memory}\label{secLSTM}
RNNs have gained significant popularity in the fields of time-series analysis and sequence modeling due to their unique ability to incorporate information from past observations through the utilization of directed cycles. This characteristic enables RNNs to effectively process current data by leveraging the knowledge acquired from previous data points. They employ a memory mechanism to remember the state of previous data but they face Vanishing and Exploding Gradient problems when learning long-term dependencies. The LSTM and GRU networks are advanced models of the RNNs, proposed to solve these problems \cite{srivastava2018comparative,ko2020deep, zhang2021deep}. 

\begin{figure}
	\centering
	\includegraphics[width=4.5 in]{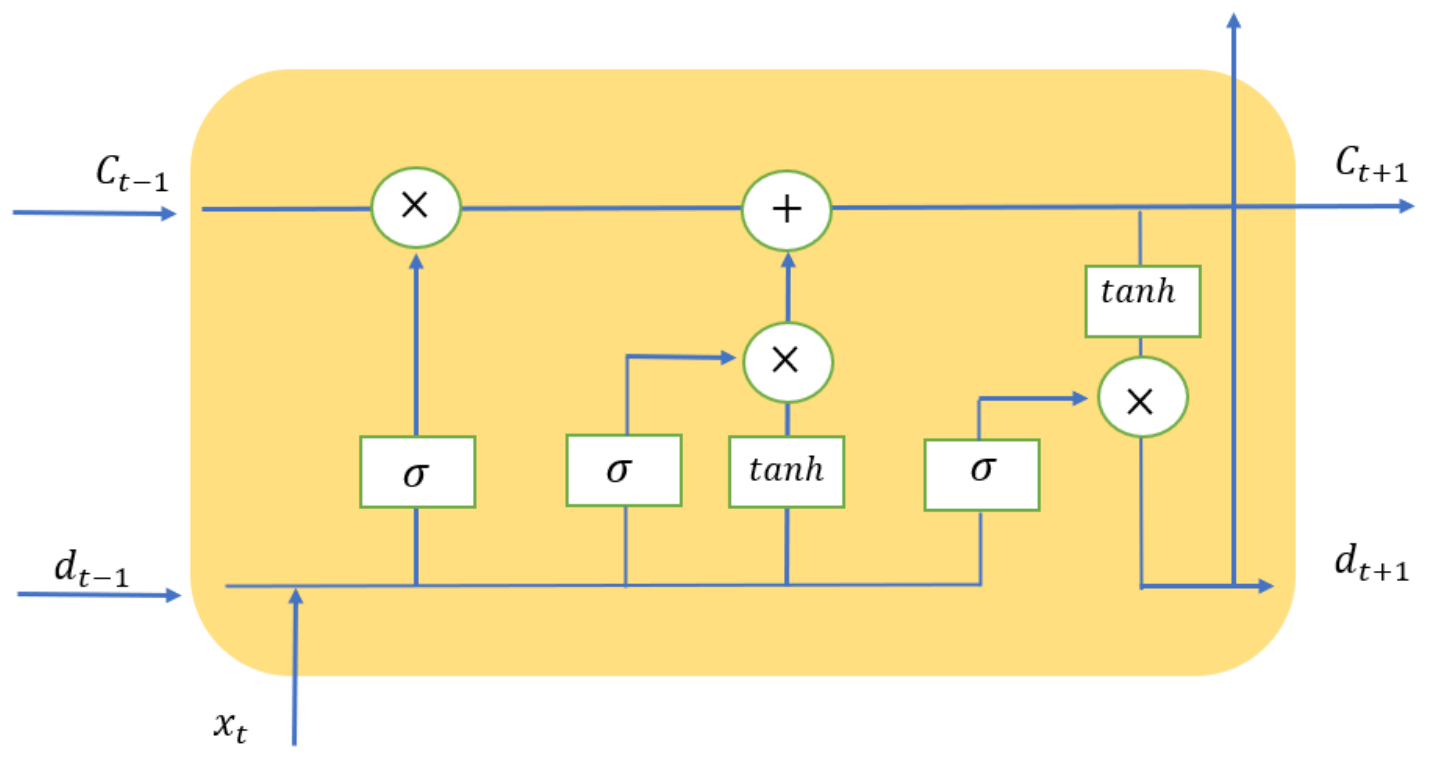}
	\caption{The LSTM structure}
	\label{2}
\end{figure}

LSTM, the fundamental information-processing units are referred to as "cells." These cells exhibit a higher level of sophistication compared to the neurons found in MLP architectures. LSTM cells, akin to neurons, can be both connected and stacked together to facilitate the transmission of temporal information. One key characteristic of LSTM is its ability to transform information into a cellular state, a feature commonly known as a "gate." The structure of the LSTM network is illustrated in Figure \ref{2}. Each LSTM unit encompasses three gates: the input gate, the forget gate, and the output gate. These gates are responsible for enabling read, write, and reset functions, respectively. The cell mode, which represents the information flow path that ensures sequential information transfer, relies on these gates for updating or discarding historical information. This ability allows the LSTM to determine the long-term relevance of different information inputs. In the LSTM equations, $C_{t-1}$ represents the cell state from the previous module, $d_{t-1}$ 1 denotes the output of the preceding module, and $X_t$ corresponds to the current input utilized for generating new memory. The output information includes the subsequent transmission of the cell state $C_t$ and the newly produced output $d_t$.

The forget gate in LSTM acts as a valve to regulate the flow of information. When the input gate remains constantly open, a large amount of information inundates the memory. To address this, a forgetting mechanism is employed, which is referred to as the forget gate. This gate examines the previous output  $d_{t-1}$ and the current input  $X_t$ and outputs a number between 0 and 1 for each digit in the cell state  $C_{t-1}$. A value of 1 indicates complete preservation, while a value of 0 signifies complete removal. The calculation formula for the forget gate is presented in Equation 3, where $W_f$ represents the weight matrix, $b_f$ denotes the bias term, and F represents the output of the network, which lies in the range (0, 1) and signifies the probability of forgetting the previous cell state. A value of 1 denotes ‘‘complete preservation‘‘, while a value of 0 signifies ‘‘complete elimination‘‘.
\begin{equation}
f_t=sigmoid(W_f[d_{t-1},X_t] + b_f)
\end{equation}

In LSTM, once the recurrent neural network "forgets" a portion of the previous state, the input gate plays a crucial role in incorporating the latest memory from the current input. This process is accomplished through the ‘‘input gate’’. The input gate in LSTM comprises two components. The first component is a sigmoid layer referred to as the ‘‘input threshold layer’’ which determines the values that need to be updated. The second component is a  $tanh$ layer, that generates a new candidate vector $\tilde{C}_t$, which is subsequently incorporated into the state. This relationship is illustrated in Equations 4, 5, and 6
\begin{equation}
h_t=\sigma (W_n.[d_{t-1},X_t] + b_n)
\end{equation}
\begin{equation}
\tilde{C}_t=\tanh (W_m.[d_{t-1},X_t] + b_m)
\end{equation}
\begin{equation}
C_t=F_t*C_{t-1} +h_t*\tilde{C}_t
\end{equation}
In Equations 4, 5, and 6, $W_n$ denotes the weight matrix, $b_n$ denotes the bias term, $W_m$ corresponds to the weight matrix used to update the unit status, $b_m$ represents the bias term used to update the unit status \cite{wang2019forecasting}, and $C_t$ denotes the updated memory unit status. Equation 7 involves the input gate  $h_t$ and $\title{C}_t$, which undergo a dot product operation to determine whether the state of the time step memory unit should be updated. The forget gate $F_t$ performs a scalar product with  $C_{t-1}$ to determine whether it is necessary to retain the initial state of the memory unit for the time step.  

The output gate in LSTM pertains to the current time output that needs to be generated after computing the new state. It also serves the purpose of controlling the level of filtering applied to the memory unit's state within this layer. The calculation formulas for the output gate are presented in Equations 7 and 8.  Firstly, the sigmoid activation function is applied to obtain  $O_t$  which falls within the range of  [0, 1]. Then, the state of the memory cell  $C_t$ is multiplied by the tanh activation function and further multiplied by  $O_t$, This resulting value represents the output of this layer. The output $d_t$ is not only influenced by the input $X_t$ at time step  $t$ and the activation value  $d_{t-1}$ from the previously hidden layer, but it also depends on the state of the memory unit  $C_t$ at the respective time step.
\begin{equation}
d_t=O_t*\tanh (C_t)
\end{equation}
\begin{equation}
O_t=\sigma (W_0.[d_{t-1},X_t] + b_0)
\end{equation}
\subsection{Hyperparameter Tuning}\label{secBasicHT}
As hyperparameter values play a critical role in the performance of deep learning networks, researchers frequently resort to techniques such as Grid and Random search to identify the most suitable hyperparameter configurations for deep neural networks\cite{alizadeh2020network}. The former is a traditional method that needs a set of manually specified values for each hyper-parameter. Then, it simply explores all the search space to find the optimal values. Even though it can be parallelized easily, the grid search suffers from the curse of dimensionality. On the other hand, a random search can sample the important dimension of search space, so performs more efficiently than a grid search. Employing nature-inspired meta-heuristic algorithms like Genetic Algorithms and Swarm Intelligence is an alternative approach to optimize hyper-parameters of neural networks with promising results. In recent times, metaheuristic algorithms have gained attention for refining hyperparameters\cite{alizadeh2021combination}. In this study, using the combination of the gray wolf optimization algorithm (GWO) and the whale optimization algorithm (WOA) in the form of a random switch is proposed to determine the optimal values of the hyperparameters. The following subsection presents a detailed explanation of this algorithm.

\subsubsection{Grey wolf optimization algorithm}
The Grey Wolf Optimizer (GWO) is a metaheuristic optimization algorithm inspired by the hunting behavior of grey wolves. Proposed by \cite{mirjalili2014grey}, GWO mimics the social hierarchy within wolf packs to solve optimization problems.
The GWO algorithm involves the following steps:\\
1. Population initialization: Set the population size and other parameters, and randomly initialize the population of wolves.\\
2. Social hierarchy: Identify the three wolves with the best fitness as alpha, beta, and delta, representing the leaders. The remaining wolves are considered omega. The optimization process is led by alpha, beta, and delta, while the omega wolves follow their lead.\\
3. Encircling prey: Wolves mimic the behavior of surrounding prey during hunting. The position update equations for a wolf are defined as:
Distance calculation: $\overrightarrow{D} = |\overrightarrow{C} \cdot \overrightarrow{X}_*(t) - \overrightarrow{X}(t)|$
Position update: $\overrightarrow{X}(t+1) = \overrightarrow{X}_*(t) - \overrightarrow{A} \cdot \overrightarrow{D}$
The coefficient vectors $\overrightarrow{A}$ and $\overrightarrow{C}$ are determined using random vectors and a decreasing linear coefficient.\\
4. Hunting: Wolves adjust their positions based on the positions of alpha, beta, and delta, assuming that they know the probable location of prey. The position update equations for a wolf are:
Distance calculation: $\overrightarrow{D}_\alpha = |\overrightarrow{C}1 \cdot \overrightarrow{D}\alpha - \overrightarrow{X}|$, and similarly for beta and delta.
Position update: $\overrightarrow{X}1 = \overrightarrow{X}\alpha - \overrightarrow{A}1 \cdot \overrightarrow{D}\alpha$, and similarly for beta and delta.
New position: $\overrightarrow{X}(t+1) = \frac{\overrightarrow{X}_1 + \overrightarrow{X}_2 + \overrightarrow{X}_3}{3}$ \\
5. Attacking prey (exploitation): Wolves continue attacking prey until it stops moving. This is modeled by changing the value of $\overrightarrow{A}$ to make the next possible location of a wolf be a combination of its current position and the prey's position.
6. Search for prey (exploration): Wolves utilize the positions of alpha, beta, and delta to search for new prey by adjusting the value of $\overrightarrow{A}$. A larger magnitude of $\overrightarrow{A}$ encourages exploration.
By simulating the social hierarchy and hunting behavior of wolves, the GWO algorithm aims to find optimal solutions to optimization problems.\\

\subsubsection{Whale optimization algorithm }
The Whale Optimization Algorithm (WOA) draws inspiration from the Bubble-net hunting strategy of whales \cite{mirjalili2016whale}. WOA comprises three main steps: encircling, exploitation, and exploration.

\textbf{Encircling}: Whales encircle potential targets using equations:

\begin{equation}
\overrightarrow{D} = |\overrightarrow{C} \cdot \overrightarrow{X}^*(t) - \overrightarrow{X}(t)|
\end{equation}

\begin{equation}
\overrightarrow{X}(t+1) = \overrightarrow{X}^*(t) - \overrightarrow{A} \cdot \overrightarrow{D}
\end{equation}

\textbf{Exploitation}: This phase involves two methods, shrinking encircling and spiral updating position, determined by a random choice governed by \(p\):

\begin{equation}
\overrightarrow{A} = 2\overrightarrow{a} \cdot \overrightarrow{r} - \overrightarrow{a}
\end{equation}

\begin{equation}
\overrightarrow{C} = 2 \cdot \overrightarrow{r}
\end{equation}

\begin{equation}
\overrightarrow{X}(t+1) =
\begin{cases} 
\overrightarrow{X}^*(t) - \overrightarrow{A} \cdot \overrightarrow{D},  & \text{if } p < 0.5 \\
\overrightarrow{D}^\prime \cdot e^{bl} \cdot \cos(2 \pi l) + \overrightarrow{X}^*(t), & \text{if } p \ge 0.5 
\end{cases}
\end{equation}

\textbf{Exploration}: This phase, with \(|\overrightarrow{A}| > 1\), relies on random agent selection using equations:

\begin{equation}
\overrightarrow{D} = |\overrightarrow{C} \cdot \overrightarrow{X}_{\text{rand}} - \overrightarrow{X}|
\end{equation}

\begin{equation}
\overrightarrow{X}(t+1) = \overrightarrow{X}_{\text{rand}} - \overrightarrow{A} \cdot \overrightarrow{D}
\end{equation}

In practice, WOA starts with random solutions, with \(\overrightarrow{a}\) decreasing from 2 to 0, supporting exploration and exploitation. Depending on \(|\overrightarrow{A}|\), agents may choose circular or spiral motion patterns. Termination conditions signal the end of the WOA algorithm.

\subsubsection{ Proposed Hybrid Metaheuristic Optimization Algorithm }
\IncMargin{1em}
\begin{algorithm2e}[t]
\SetAlgoLined
\caption{RS-GWO-WOA Algorithm}
\label{alg:RS-GWO-WOA}
\SetKwInOut{Input}{input}\SetKwInOut{Output}{output}
\BlankLine
\Input{Population size $n$, Number of iterations, Initialization parameters}
\Output{Optimal solution $X_*$}
\BlankLine
$Initialize$ population of $n$ agents $X_i\ (i=1,2,…,n)$ \;
$Initialize$ parameters $a, A, C, l,$ and $p$\;
Evaluate the fitness of all agents by calculating the fitness of their locations\;
$X_\alpha$ = best agent\;
$X_\beta$ = second-best agent\;
$X_\delta$ = third-best agent\;
\BlankLine
\While{$t \leq$ number of iterations}{
    Generate a random number $\varepsilon \in [0,1]$\;
    \uIf{$\varepsilon < 0.5$}{%
        \ForEach{whale}{%
            Update $a, A, C, l,$ and $p$\;
            \uIf{$p < 0.5$}{%
                \uIf{$|A| < 1$}{%
                    Update the position of the current whale using Eq. (16)\;
                }
                \uElseIf{$|A| \geq 1$}{%
                    Select a random whale ($X_{\text{rand}}$)\;
                    Update the position of the current whale using Eq. (23)\;
                }
            }
            \uElseIf{$p \geq 0.5$}{%
                Update the position of the whale using Eq. (20)\;
            }
        }
        Check if any whale goes beyond the search space and adjust its position\;
        Calculate the fitness of each whale\;
        Update $X_*$ if a better solution is found\;
    }
    \uElseIf{$\varepsilon \geq 0.5$}{
        \ForEach{wolf $X_i$ in the population}{%
            Update the position of the current wolf using Eq. (15)\;
        }
        Update $a, A, C, l$\;
        Evaluate the fitness of all wolves in the population\;
        Update $X_\alpha, X_\beta,$ and $X_\delta$\;
    }
    $t = t + 1$\;
}
\textbf{return} $X_*$
\end{algorithm2e}

The hybrid GWO-WOA algorithm, a fusion of the Grey Wolf Optimizer (GWO) and the Whale Optimization Algorithm (WOA), presents a powerful approach for addressing optimization problems with improved search capabilities and solution quality. By synergistically harnessing the exploration prowess of WOA and the exploitation capabilities of GWO, this hybrid algorithm strives to strike an optimal balance between exploration and exploitation, resulting in enhanced performance in solving complex optimization problems.\\

The algorithm entails a series of steps to leverage the strengths of both GWO and WOA in a cohesive manner. It begins with an initialization phase, where the population size and other pertinent parameters are set. The algorithm then proceeds to the GWO exploration phase, which establishes the social hierarchy among the wolves by identifying the alpha ($\alpha$), beta ($\beta$), and delta ($\delta$) wolves based on their fitness. The remaining wolves are designated as omega ($\omega$) wolves, forming a hierarchical structure.\\
During the GWO exploration phase, the omega wolves dynamically update their positions using the encircling prey equation. This equation incorporates the position of the prey, represented by $\overrightarrow{X}_*(t)$, and a coefficient vector $\overrightarrow{A}$ to calculate the new positions $\overrightarrow{X}(t+1)$ of the omega wolves. The distance $\overrightarrow{D}$ between each wolf and the prey guides their movement, ensuring a convergent trajectory toward optimal solutions.\\

Following the GWO exploration phase, the algorithm transitions to the WOA exploitation phase. In this phase, the positions of the whales are updated using exploration equations based on the positions of the target, represented by $\overrightarrow{X}{\text{target}}$, and a random vector $\overrightarrow{C}{\text{rand}}$. These updates allow the whales to efficiently explore the search space while exploiting promising regions.\\

To strike a balance between exploration and exploitation, a careful adjustment of the position updates is performed using a balance factor. This factor modulates the impact of exploration and exploitation, ensuring that the algorithm maintains a fine equilibrium throughout the optimization process.\\
Fitness evaluation is conducted to assess the quality of solutions generated by the algorithm. The fitness values of both wolves and whales are calculated based on the objective function of the optimization problem at hand. Subsequently, the alpha, beta, and delta leaders are selected based on their fitness values, reflecting the most promising individuals within the population.\\
The algorithm iteratively repeats the GWO exploration phase and the WOA exploitation phase until a termination criterion is met. This criterion can be either reaching the maximum number of iterations or obtaining a satisfactory solution. By cyclically engaging in exploration and exploitation, the hybrid GWO-WOA algorithm effectively harnesses the strengths of both algorithms, leading to improved search capabilities and high-quality solutions. It is important to note that the specific equations and parameters used in GWO and WOA can be tailored to the characteristics of the optimization problem under consideration. Adapting these equations and parameters accordingly enables the algorithm to effectively address diverse problem domains and achieve superior optimization performance.\\

\section{Forecasting Framework}\label{SecForecasting}
The proposed method combines the CNN and LSTM networks for forecasting the number of infected, cured, and dead people. As described in \ref{secCNN}, the CNNs are able to extract local and deep features using their convolutional layers. On the other hand, the LSTM models can capture long-term dependencies, making them a good choice for sequence modeling. As a result, their combination can improve the accuracy of prediction \cite{guo2020short}. \\
\IncMargin{1em}
\begin{algorithm2e}
\SetAlgoLined
\caption{Hybrid CNN-LSTM Algorithm with Hyperparameter Tuning}
\label{alg:hybrid_cnn_lstm}
\SetKwInOut{Input}{input}\SetKwInOut{Output}{output}
\BlankLine
\Input{Input data sequence $\mathbf{X}$, CNN parameters, LSTM parameters}
\Output{Forecasted values}
\BlankLine
$\mathbf{X}_{\text{cnn}}$ = Apply CNN to $\mathbf{X}$\;
$\mathbf{X}_{\text{cnn}}$ = Reshape $\mathbf{X}_{\text{cnn}}$ to match LSTM input shape\;
Initialize LSTM model with parameters: batch size, learning rate, hidden units, etc.\;
Train LSTM model on $\mathbf{X}_{\text{cnn}}$ with hyperparameter tuning\;
\BlankLine
\For{$i$ in range(number of forecast steps)}{
    $\mathbf{X}_{\text{forecast}}$ = Get last input sequence from $\mathbf{X}_{\text{cnn}}$\;
    $\mathbf{X}_{\text{forecast}}$ = Reshape $\mathbf{X}_{\text{forecast}}$ to match LSTM input shape\;
    $\mathbf{y}_{\text{forecast}}$ = LSTM model.predict($\mathbf{X}_{\text{forecast}}$)\;
    Append $\mathbf{y}_{\text{forecast}}$ to $\mathbf{X}_{\text{cnn}}$\;
}
\BlankLine
\textbf{return} Forecasted values from $\mathbf{X}_{\text{cnn}}$
\end{algorithm2e}\\
Consequently, this research endeavor integrates the CNN network with the LSTM network, capitalizing on their individual strengths. To achieve this, a parallel network connection is employed to construct the CNN-LSTM network model, effectively leveraging temporal and spatial characteristics from both networks. The comprehensive structure of the proposed approach is illustrated in  Figure \ref{figOveall}. Firstly, the dataset undergoes reprocessing, after which the preprocessed data is fed into the Forecasting Engine module. Subsequently, the feature information extracted from the CNN and LSTM networks is separately processed in the same dimension through the map layer. Furthermore, the outputs of the CNN and LSTM networks are concatenated through a parallel connection. Finally, the classification task is performed using an activation function in the output layer. Additionally, the RS-GWO-WOA optimization algorithm is employed to fine-tune the hyperparameters of the CNN-LSTM model. The overall architecture of the proposed approach is presented in Figure 3. Each of these steps will be elucidated in greater detail in the subsequent sections.

\subsection{Preprocessing}

Preprocessing involves cleansing and normalizing the dataset. To cleanse the dataset, all missing and defective values are replaced by the average value of the next day and the previous day of their respective column. Next, the values are normalized by the Min-Max Scaler method of Keras.

\subsection{Forecasting Engine}
As shown in Figure \ref{figOveall}, this module is composed of a combination of CNN and LSTM. Firstly, the preprocessed input data is fed into the convolutional layer. The convolutional layer extracts the local features and makes the feature maps. The pooling layer receives the feature maps, reduces the dimensions, and finds the essential features. Indeed, this layer decreases the dimension and computation time. The proposed method utilizes Max Pooling which is common in dimension reduction.

The flattened layer receives the output of the pooling layer, and flats the data. Later on, the Repeat vector layer changes the dimension of data to be compatible with the LSTM layer. Then, the LSTM layer is utilized to capture the long-term dependencies. The output of LSTM is fed into the Dense layer to prepare the output.
\begin{figure}[H]
	\centering
	\includegraphics [width=6.2 in] {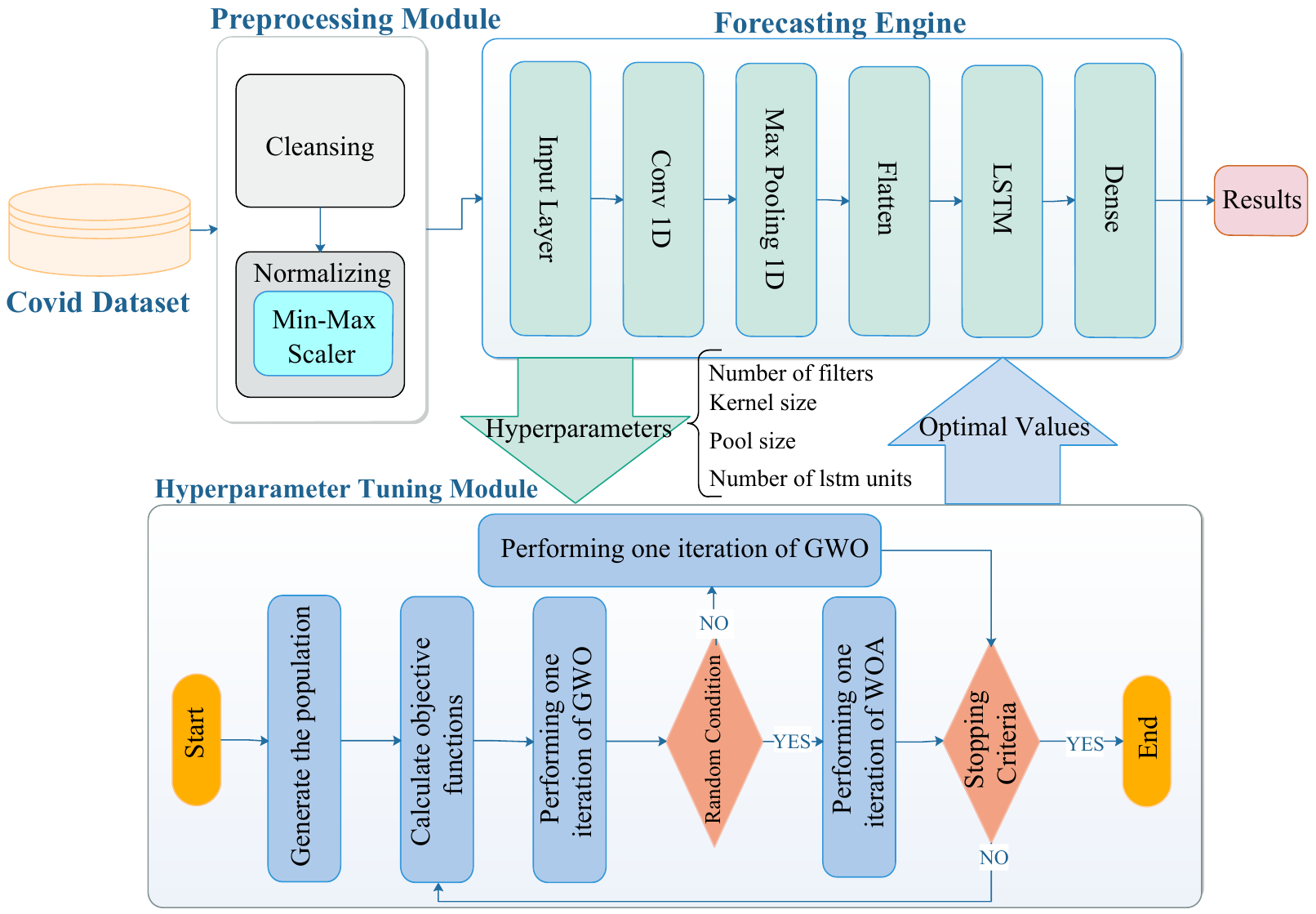}
	\caption{Overall structure of the proposed method}
	\label{figOveall}
\end{figure}
\subsection{Hyperparameter Tuning Module}\label{secProHTM}
As described in Section \ref{secBasicHT}, the performance of deep learning methods is highly dependent on the values of hyperparameters. Therefore, the proposed method uses the GWO-WOA algorithm to fine-tune the hyperparameters of the CNN-LSTM model such as the number of filters and kernel size of the convolutional layer, the pool size of the Max-pooling layer, and the number of units of the LSTM layer. We run the proposed method with the data related to each country individually and find the optimal values of hyperparameters by the GWO-WOA.\\
GWO-WOA algorithm takes advantage of GWO and WOA by integrating them in two ways. WOA updates the positions of agents according to a randomly generated solution to avoid being trapped in locally optimal solutions. In comparison, GWO updates the position of each solution based on the positions of the three best solutions in the population, and this strategy makes the GWO more exploitation-oriented than the WOA. The balance between exploration and exploitation plays an important role in improving the performance of a meta-heuristic algorithm. Hence, to obtain this balance and take advantage of both, we combine these two algorithms in random switching. Therefore, there will be a new version of GWO-WOA:  Random Switcher GWO-WOA.\\ 
Random Switcher GWO-WOA (RS-GWO-WOA) randomly switches between GWA and WOA in each iteration, so the computational time of finding the optimal solution decreases. For instance, $a$ random number between 0 and 1 is generated, if the random number is greater than 0.5, the GWO algorithm is performed, otherwise, the WOA is selected. In this approach, control of operator selection from GWO or WOA is performed using a random factor without considering the performance information of each algorithm. Hence, one of them may not be able to improve the population and get stuck in the Local Optima (LO), and it may not be possible to find better solutions. The pseudo-code of RS-GWO-WOA is shown in algorithm 1. It is noteworthy that the updating of parameters a and $\xi$ can help in shifting from exploration to exploitation trends and avoid getting trapped in local optimums. 

\section{Results and Discussion}\label{secResults}
In this section, we evaluate the proposed method with the real-world dataset, described in Section \ref{secDataDescription}. Additionally, we compare it with other state-of-the-art methods such as CNN, RCLSTM, CNN-GRU, BILSTM, and ConvLSTM. This section includes Evaluation Setup and metrics, Hyperparameter tuning, and Comparison with other methods. 
\section{Data Description}

This study focuses on examining the trajectory of COVID-19 growth in countries characterized by the highest infection rates of the disease. In general, twenty-four countries from six continents have been studied. Figure 4 shows the growth trend of this disease in these twenty-four countries.
\begin{figure}[H]
	\centering
	\includegraphics[width=5.6 in]{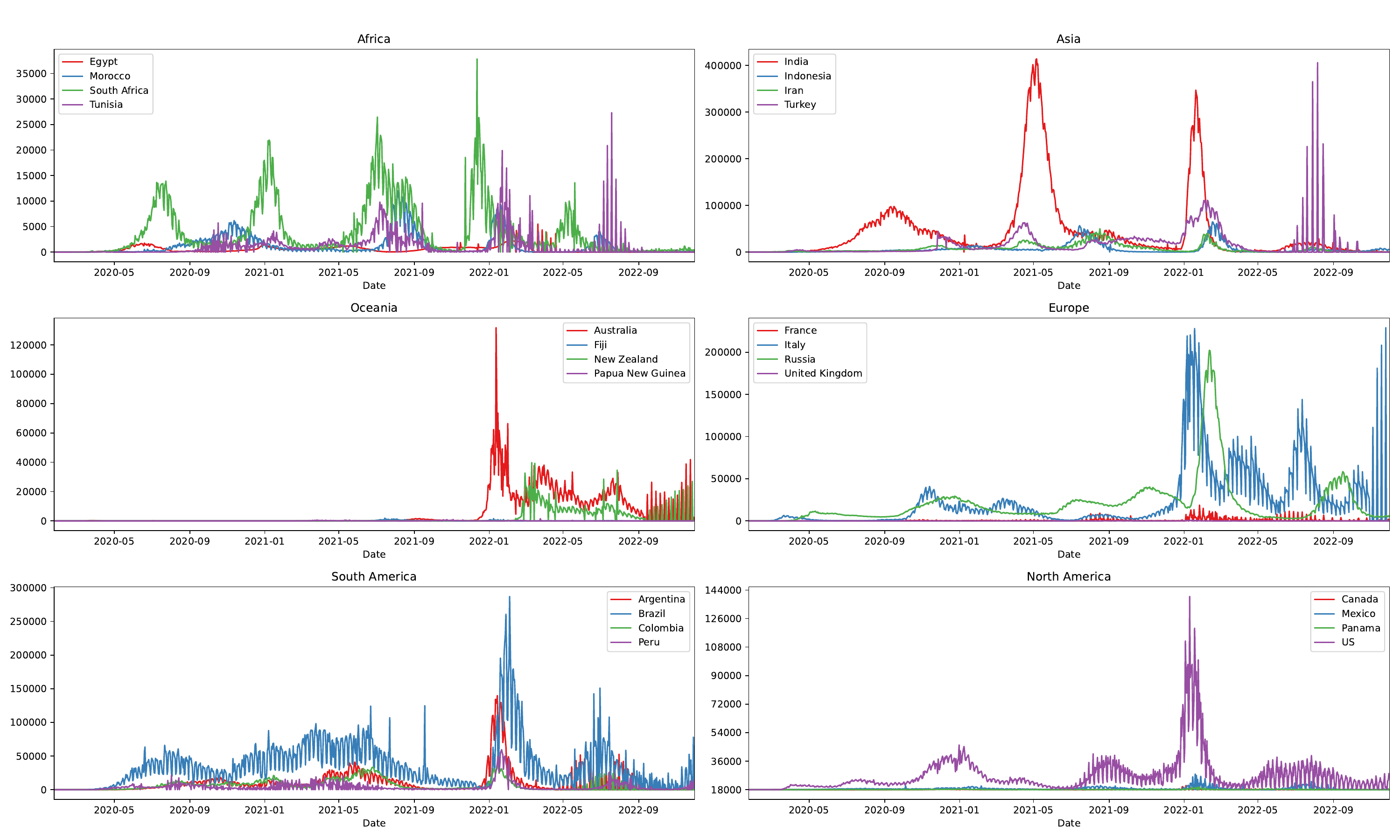}
	\caption{The growth trend of confirmed COVID-19 infected patients in 24 countries by continent}
	\label{FIG:1}
\end{figure}
\begin{figure}[H]
	\centering
	\includegraphics[width=5.6 in]{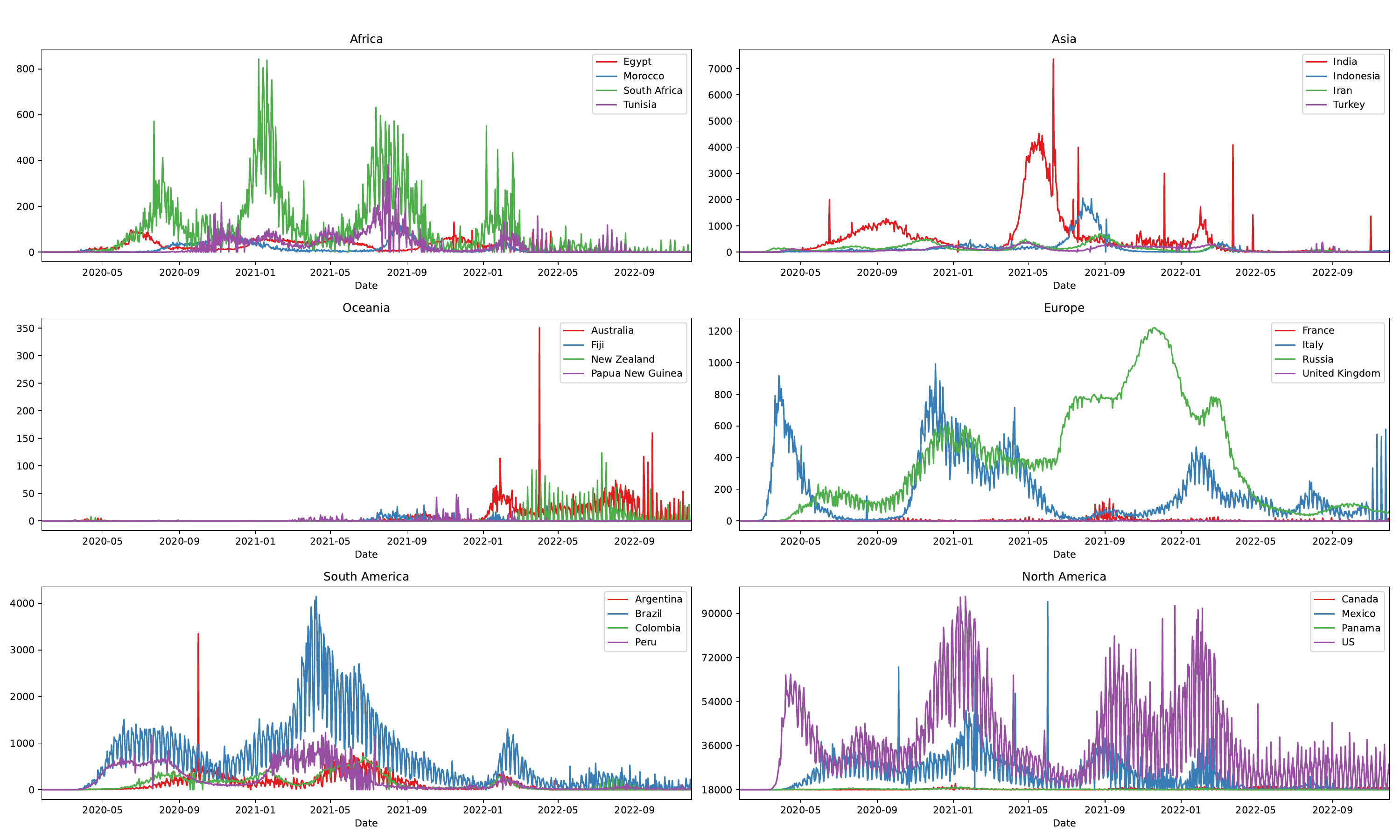}
	\caption{The growth trend of confirmed COVID-19 infected patients in 24 countries by continent}
	\label{FIG:1}
\end{figure}

The dataset used in this article is time series and multivariate. The variables used are the number of people infected with corona disease, the number of people who died, and the number of people who recovered from 3/22/2020 to 12/1/2022, collected daily. The proposed model has been implemented for all variables for all countries to predict the number of people who have died, the number of people who have recovered, and the number of people with coronavirus disease for all countries. Also, the first 80\% of the data is used for training the model, and the last 20\% of the data is used for testing and evaluating the model.

In general, the data from twenty-four countries from six continents have been used to evaluate and check the accuracy of the proposed model. The countries surveyed on the American continent are Canada, the United States, Argentina, Brazil, Mexico, Peru, Panama, and Colombia. European countries studied include Russia, France, Italy, New Zealand, and the United Kingdom. The four countries with the highest rates of COVID-19 in Asia are Iran, India, Turkey, and Indonesia. Selected countries from the African continent are Egypt, Morocco, Africa, and Tunisia. Also, the COVID-19 data in New Guinea, Australia, and Fiji from Australia and Oceania were obtained. In the next step, we examined the growth and decline of the number of people with coronary heart disease among these twenty-four countries to use deep learning to predict the number of patients, the number of people who have recovered, and the number of people who have died from coronary artery disease. The twenty-four countries examined in this study are shown in Figure \ref{FIG:1}.\\

\subsection{Evaluation Setup and Metrics}
The proposed methodology was implemented using Python version 3.8 on the Google Colab \footnote{https://colab.research.google.com/} platform, equipped with a K80 GPU and 12 GB of RAM. The implementation utilized the NiaPy \cite{NiaPyJOSS2018} and Keras \cite{chollet2015keras} libraries for incorporating the metaheuristic algorithms and deep learning models. To account for the stochastic nature of deep learning networks and their tendency to yield varying results with each run, the algorithms were executed ten times, and the average outcomes were subsequently reported. The batch size was set to 1, and the number of epochs was set to 100.

As previously mentioned, the evaluation of performance and predictive effects was conducted using MAE and MSE, and R-square ($R^2$) metrics. The calculation formula for MAE, as depicted in Equation 16, involves the predicted value  $y_i$ and the true value $x_i$. 
\begin{equation}
MAE=  \frac{1}{n} \sum_{i=1}^n|y_i-x_i |
\end{equation}
Similarly, the calculation formula for Mean Square Error (MSE), as presented in Equation 17, involves the predicted value $y_i$ and the true value  $x_i$. 
\begin{equation}
MSE=  \frac{1}{n} \sum_{i=1}^n(y_i-x_i)^2
\end{equation}
\begin{table*}[!htb]
\caption{Configuration settings of metaheuristic algorithms}
\begin{tabular}{|c|c|c|}
\hline
\textbf{Algorithm}            & \multicolumn{1}{c|}{\textbf{Hyperparameter}} & \multicolumn{1}{c|}{\textbf{Value}} \\ \hline
\multirow{3}{*}{\textbf{GWO}} & $\overrightarrow{a}$         & Linearly   decreased from 2 to 0    \\
                              & $\overrightarrow{r}_1$      & A random   vector in {[}0,1{]}      \\
                              & $\overrightarrow{r}_2$       & A random   vector in {[}0,1{]}      \\ \hline
\multirow{2}{*}{\textbf{WOA}} & $\overrightarrow{a}$          & Linearly   decreased from 2 to 0    \\
                              & $\overrightarrow{r}$          & A random   vector in {[}0,1{]}      \\ \hline
\multirow{4}{*}{\textbf{GA}}  & Crossover                                    & Uniform   crossover                 \\
                              & Mutation                                     & Uniform   mutation                  \\
                              & Crossover rate                               & 0.25                                \\
                              & Mutation rate                                & 0.25                                \\ \hline
\end{tabular}
\end{table*}
\subsection{Hyperparameter Tuning}
As indicated in Section  \ref{secProHTM}, the proposed approach employs the GWO-WOA to optimize the hyperparameters of the CNN-LSTM model. These hyperparameters include the number of filters, kernel size of the convolutional layer, pool size of the Max-pooling layer, and the number of units in the LSTM layer. Furthermore, a comparative analysis was conducted to evaluate the performance of GWO-WOA in comparison to other optimization algorithms, namely Genetic Algorithm (GA), GWO \cite{mirjalili2014grey}, and WOA \cite{mirjalili2016whale}. The configuration settings for these algorithms, as proposed in the NiaPy library, are presented in Table 1.\\
We fine-tune the hyperparameters of the CNN-LSTM model for each country, which includes the number of filters (candidate values: 32 and 64), kernel size (candidate values: 3, 4, 5, 6, 7, and 8), pool size (candidate values: 2, 3, and 4), and the number of LSTM units (candidate values: 10, 15, 20, and 25).\\

\subsection{Comparison With Other Methods}
In this section, a comparison is made between the results of the designed method and other approaches, namely CNN, RCLSTM, BiLSTM, CNN-GRU, and ConvLSTM. The objective is to evaluate and compare the performance of these methods in order to assess the effectiveness and superiority of the proposed approach.
\begin{table*}[h]
\caption{Advantages and Disadvantages of Deep Learning Methods for COVID-19 Forecasting}
\centering
\begin{tabular}{lp{6cm}p{6cm}}
\hline
Method & Advantages & Disadvantages \\
\hline
CNN & 
- Effective in capturing spatial patterns in data  & 
- May overlook temporal dependencies   \\
& - Able to learn local features & 
- Limited ability to capture long-term dependencies \cite{han2022survey} \\
& - Fast training and prediction times & \\
\hline
RCLSTM & 
- Captures both spatial and temporal dependencies & 
- More complex architecture \\
& - Able to model long-term dependencies & 
- Longer training times \\
& - Can handle irregular time intervals & \\
\hline
ConvLSTM & 
- Captures both spatial and temporal dependencies (Reference: ) & 
- Higher computational complexity compared to individual models \\
& - Effective in modeling sequential data with spatial relationships & 
- Requires more training data\cite{rahman2019deep} \\
& - Can handle variable-length sequences & \\
\hline
CNN-GRU & 
- Efficient in processing sequential data  & 
- May struggle with capturing long-term dependencies \\
& - Less prone to overfitting & 
- Less expressive than LSTM \cite{shah2021deep} \\
& - Faster training times compared to LSTM & \\
\hline
BILSTM & 
- Able to capture dependencies in both forward and backward directions & 
- Slower training and inference times \\
& - Effective in modeling sequential data & 
- Higher memory requirements\cite{wozniak2023bilstm} \\
& - Suitable for tasks with bidirectional dependencies & \\
\hline
CNN-LSTM (Hybrid) & 
- Combines strengths of CNN and LSTM   & 
- Relatively higher computational complexity \\
& - Effective in capturing spatial and temporal dependencies & 
- May require more training data\cite{lindemann2021survey} \\
& - Able to model both local and global features & \\
\hline
\end{tabular}
\label{tab:advantages_disadvantages}
\end{table*}

Each method has its own strengths and limitations. CNNs are effective in capturing spatial patterns in the data and learning local features rapidly, making them suitable for spatial analysis. However, they may overlook temporal dependencies and struggle to capture long-term relationships in the data.\\
RCLSTMs, on the other hand, excel in capturing both spatial and temporal dependencies, allowing for the modeling of long-term relationships and handling irregular time intervals. They are capable of learning complex patterns; however, their architecture is more complex, leading to longer training times.\\
The hybrid CNN-LSTM approach combines the strengths of CNNs and LSTMs, making it a powerful choice for COVID-19 forecasting. It effectively captures both spatial and temporal dependencies, allowing for the modeling of local and global features. While it offers superior performance.\\
CNN-GRU models are efficient in processing sequential data, have faster training times compared to LSTMs, and are less prone to overfitting. However, they may struggle with capturing long-term dependencies and are generally less expressive than LSTMs.\\
BILSTMs, which capture dependencies in both forward and backward directions, are effective in modeling sequential data and are suitable for tasks with bidirectional dependencies. However, they have slower training and inference times and higher memory requirements.\\
ConvLSTMs capture both spatial and temporal dependencies and effectively model sequential data with spatial relationships. They can handle variable-length sequences but have higher computational complexity compared to individual models and require more training data.

In conclusion, while all the mentioned deep learning methods have their advantages and disadvantages, the hybrid CNN-LSTM approach stands out for COVID-19 forecasting. Its ability to capture both spatial and temporal dependencies, model local and global features, and strike a balance between spatial pattern recognition and capturing long-term dependencies make it a powerful choice for accurate and robust predictions. It is important to note that CNN-LSTM and ConvLSTM are distinct methods, with CNN-LSTM combining convolutional and recurrent neural networks, while ConvLSTM utilizes convolutional LSTM architecture to capture spatial-temporal information. The comparison of methods is listed in Table 2 for more clear demonstration.
\begin{figure}
	\centering
	\includegraphics [width=6.2 in] {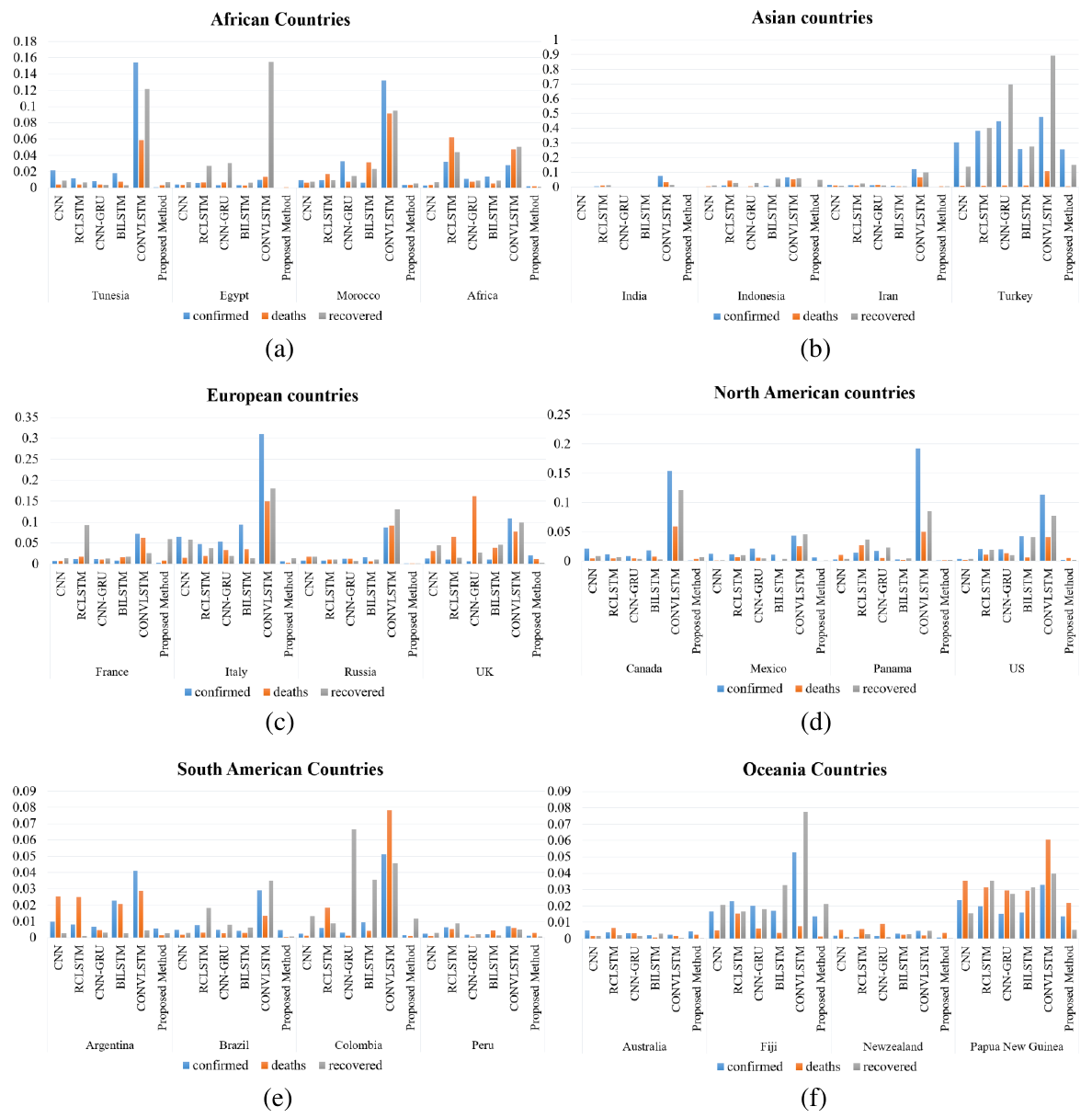}
	\caption{Comparing forecasting results}
	\label{figOveall}
\end{figure}
\subsubsection{Africa}
As shown in Figure 6 (a), in the case of infected and cured people, the proposed method, which uses the CNN-LSTM model, achieved the lowest loss in all countries compared to other models. In addition, regarding the dead people, in Egypt, Morocco, and Africa, the CNN-LSTM outperforms other methods in terms of loss, but the BiLSTM achieved a lower error in Tunisia. On the other hand, in Tunisia, the combined models performed better. Moreover, it can be seen that in Egypt, predicting the dead people has shown the lowest error, in some cases half of the other methods. Also, in Morocco, predicting the infected and cured people achieved a lower loss than the dead people. In Africa, the proposed method has reached the lowest error for predicting the infected, cured, and dead people.

\subsubsection{Asia}
Regarding Asia, we have forecasted the number of infected, cured, and dead people in India, Indonesia, Iran, and Turkey. As can be seen in Figure 6 (b), the proposed method obtained the lowest loss for forecasting the number of infected people in Indonesia, Iran, and Turkey, but in India, the CNN method showed the lowest error. In addition, the CNN-LSTM model forecasted the number of cured people in four Asian countries with the lowest errors. In the case of forecasting the dead people of Indonesia and Turkey, the CNN model achieved the lowest error. Still, the proposed method performed better than other models in India and Iran. 

Also, it can be seen that for India, the CNN-GRU model performs better in forecasting the cured and dead people, but the CNN model achieves the lowest error in forecasting the infected people. For Indonesia and Turkey, the proposed method achieved the lowest loss in predicting the number of infected and cured people, but the CNN model outperformed the other models in forecasting the number of dead people. What is more, regarding Iran, the proposed method forecasted the number of infected, cured, and dead people with the lowest error.
\begin{figure}
	\centering
	\includegraphics [width=6.2 in] {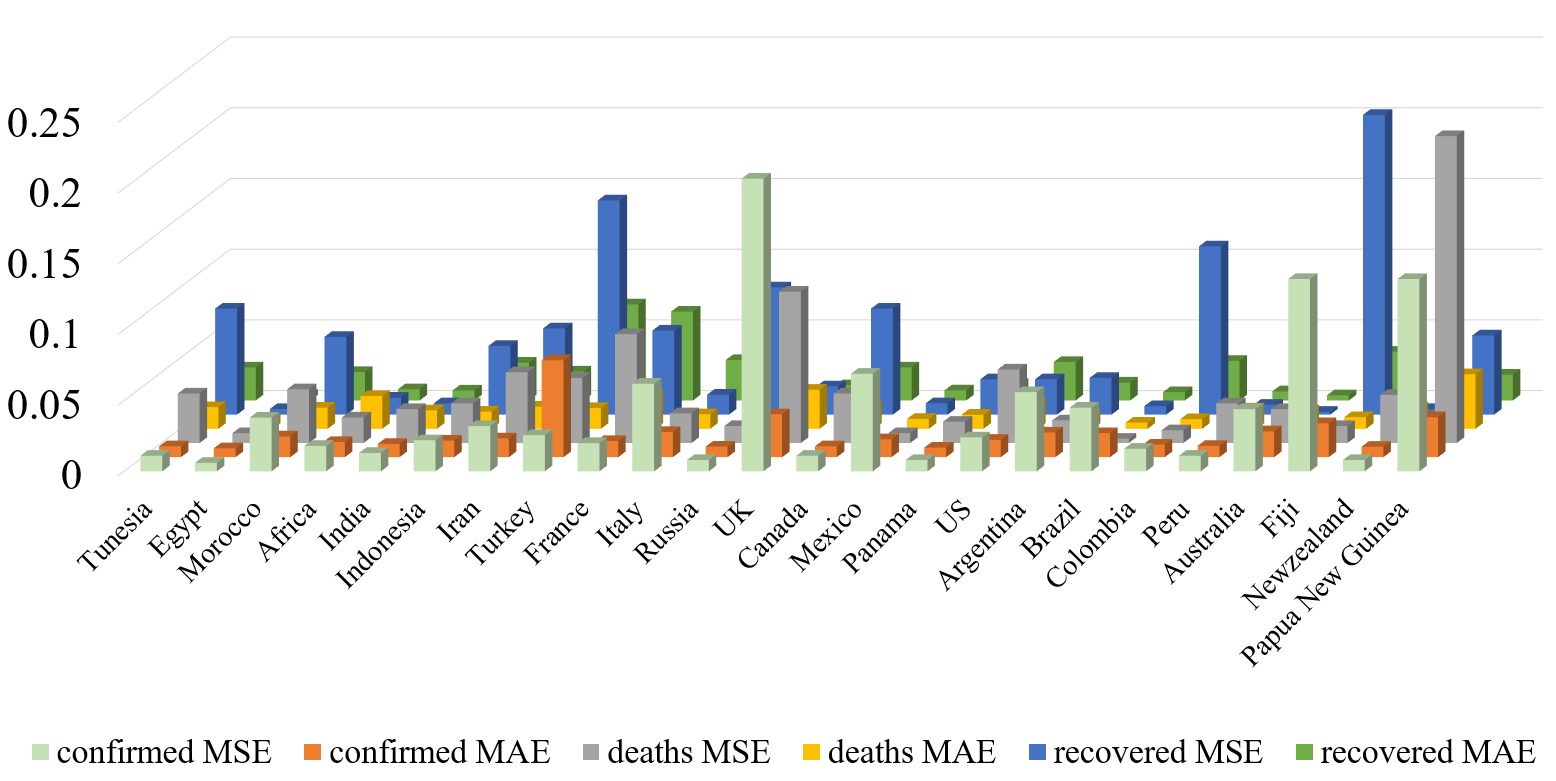}
	\caption{Comparison optimized forecasting of the proposed method in all 24 countries on six continents}
	\label{figOveall}
\end{figure}
\subsubsection{Europe}
In the context of forecasting the number of infected, cured, and deceased individuals in four European countries (France, Italy, Russia, and the UK), an analysis was conducted using the proposed method and other models including CNN, RCLSTM, BiLSTM, CNN-GRU, and ConvLSTM. The results, as presented in Figure 6 (c), demonstrate that the proposed method achieved the lowest error in forecasting the number of infected and cured individuals for Italy and Russia. For France and the UK, the CNN model outperformed the other models in predicting the number of infected individuals, while the CNN-GRU model showed better performance in forecasting the number of infected individuals for the UK.

Additionally, when predicting the number of deceased individuals, the CNN-LSTM model yielded the lowest error for Russia and the UK, while the CNN-GRU and BiLSTM models demonstrated superior performance for France and Italy, respectively. Specifically, the CNN model exhibited better forecasting results for the number of infected and cured individuals in France, while the CNN-GRU model achieved the lowest loss in predicting the number of deceased individuals. In the case of Italy, the CNN-LSTM model performed better than the other models in forecasting the number of infected and cured individuals.

\subsubsection{North America}
We have investigated the data from Canada, Mexico, Panama, and the US in America. Figure 6 (d) shows the forecasted results. As it is obvious in Figure 6 (d), for forecasting the infected and dead people, the CNN-LSTM model and the cured people with the BiLSTM model have achieved the lowest loss (0.0000804, 0.000836, 0.000066).

Regarding Canada, Panama, and the US, the CNN-LSTM outperformed the other models for forecasting the number of infected people (0.0001094, 0.0000804, 0.0002446). Also, to forecast the number of infected, cured, and dead people in Mexico, the BiLSTM method achieved the lowest loss (0.000066).  For all four countries, the CNN model achieved the lowest loss. Totally the BiLSTM model showed better performance while forecasting the cured people of Mexico (0.000066).

\subsubsection{South America}
In South America, the infected cured, and dead people of Argentina, Brazil, Colombia, and Peru are forecasted, and the results are shown in Figure 6 (e). As listed in Figure 6 (e), in the case of infected and cured people, the proposed method showed the lowest loss (0.000107, 0.000028) when forecasting the data of Peru and Brazil, respectively. Regarding the dead people, the CNN-LSTM has achieved the lowest results when forecasting the data for Brazil (0.0000556).

On the other hand, in the case of Argentina, the proposed method has achieved the lowest loss for forecasting the number of infected and cured people (0.000563, 0.0001626), but for forecasting the number of dead people, the BiLSTM model performed better (0.0000924). Also, the CNN-LSTM forecasted the number of cured and dead people in Brazil with the lowest loss (0.000028,0.0000556). The BiLSTM model achieved the lowest loss in forecasting the number of infected people in Brazil (0.0000421). For forecasting the number of infected and cured people in Colombia, the proposed method outperformed the other models (0.000159,0.000826), but the RCLSTM model performed better in the case of dead people (0.001186). Regarding the number of infected and dead people in Peru, the CNN-LSTM model has achieved the lowest loss (0.000107, 0.000072), but for forecasting the number of cured people, the CNN-GRU method outperformed the other models (0.000064). Generally, the CNN-LSTM model has achieved the best result in forecasting the number of dead people in Brazil (0.0000556).

\subsubsection{Oceania}
In Oceania, we have forecasted the number of infected, cured, and dead people in Australia, Fiji, New Zealand, and Papua New Guinea. Figure 6 (f) shows the evaluation results. 

For forecasting the number of infected and dead people, the proposed method outperformed other New Zealand and Australian models, respectively (0.000083, 0.00002425). Regarding the cured people, the BiLSTM model performed better (0.00006929) in Australia.

The BiLSTM model has achieved the lowest loss for forecasting the number of infected and cured people in Australia (0.000228,0.000069). The CNN-LSTM performed better than other methods for forecasting the number of dead people in Australia (0.000024). In the case of forecasting the number of infected and cured people of Fiji, the CNN-LSTM has achieved the lowest loss (0.00136,0.0001159), but the RCLSTM model outperformed the other models in forecasting the number of dead people (0.001676). Regarding New Zealand, the proposed method has shown the lowest loss in forecasting the number of infected people (0.000083). The ConvLSTM model outperformed the other methods for forecasting the number of infected people (0.0001974), and for forecasting the number of dead people, the CNN-GRU has shown the lowest loss (0.0001046). The proposed method has achieved the lowest loss in forecasting the number of infected, cured, and dead people (0.00136, 0.002173, 0.000552).

The proposed method generally performed better than other methods in forecasting the number of dead people in Australia (0.00002425).

\subsubsection{Comparison of continents}
In addition, we have evaluated the optimization of the proposed method on the data of six continents by the whale algorithm. As listed in Figure 7, for forecasting the number of infected, cured, and dead people, the proposed method has achieved the lowest results in Africa, South America, and North America (0.00018235, 0.00013915, 0.00033195) and the results were well-matched by the hybrid heuristic algorithm has been improved.\\

\subsection{Statistical Comparison of Different Methods}

In this section, the Friedman test and the Nemenyi post-hoc test are used for more precise evaluation to distinguish the performance of different forecasting methods. This test is non-parametric and does not hypothesize normally distributed values for the sample\cite{nemenyi1963distribution}. The null hypothesis for the conducted Friedman test concludes that all methods are equivalent and there is no significant difference between them by having similar rank.  In the Nemenyi post-hoc test if the average ranking for two different methods is greater than the critical difference their performances are considerably different. The average ranks by the Friedman test are presented in Table 3.\\

\begin{table} [b]
    \centering
    \caption{Rankings of Methods by Friedman Test}
    \begin{tabular}{lcl}
        \toprule
        Method & Rank \\
        \midrule
        CNN & 7.5 \\
        RCLSTM & 12.17 \\
        CNN-GRU & 4.15 \\
        BiLSTM & 7.0417 \\
        ConvLSTM & 8.67 \\
        CNN-LSTM(Proposed method) & 2.93 \\
        \bottomrule
    \end{tabular}
\end{table}

To calculate the Friedman statistic the following equation is used \cite{nemenyi1963distribution}:
\begin{equation}
    \left [ \frac{12}{nk(k+1)} \sum_{i=1}^k R_i^2 \right ] - 3n(k+1) = 5.5
\end{equation}

$N$ is the number of tests carried out and k and $R_i$ denote the number of methods and their ranks, respectively. By considering the significance value of 0.05, the critical value for 6 different approaches in 24 different tests(countries) is approximately 3.146. Consequently, since the Friedman statistic of  5.5  is greater than the critical value of 3.146, the null hypothesis is rejected and all methods do not have equivalent performance. Therefore, the Nemenyi test can be conducted to compare the approaches as their performance concluded to have significant differences. In order to perform the Nemenyi post-hoc test, the following equation is considered:\\

\begin{equation}
  CD= q_{\alpha}\sqrt{\frac{k(k+1)}{6N}}=2.728\sqrt{\frac{6(6+1)}{6\times24}}  = 1.474
\end{equation}
The $q_{\alpha}$ is the Nemenyi critical value for the significance value of 0.05 and 6 methods are approximately calculated as the value 2.728. The CD value of 1.474 declares that methods with an average rank difference in the range CD have the same performance. In Figure 6, the results are illustrated for better comparison.
\begin{figure}[t]
	\centering
	\includegraphics [width=4.5 in] {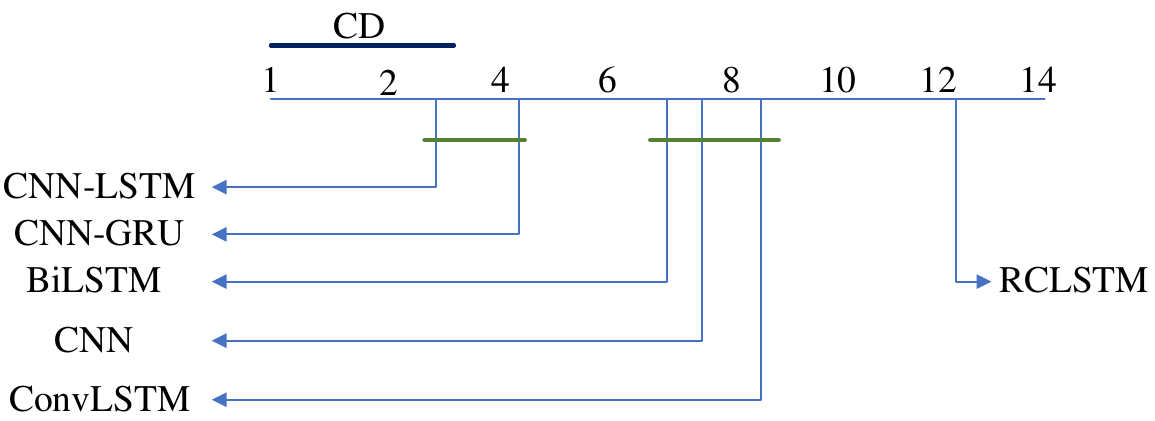}
	\caption{Visualization of statistical test (Friedman and Nemeny post-hoc) for different deep learning method performance comparison.}
	\label{figOveall}
\end{figure}
From figure 6, it can be seen that CNN-LSTM outperforms other methods. However, the difference between CNN-LSTM and CNN-GRU was not significant. Furthermore, CNN, ConvLSTM and BiLSTM have almost similar performance. RCLSTM has the worst performance overall in comparison with other methods.

\section{Conclusion}
 This study introduced a prediction model that synergistically combined a CNN and LSTM network to forecast the counts of infected, recovered, and deceased individuals across 24 countries spanning six continents. The CNN component leveraged convolutional layers to extract local and deep features while employing pooling layers for dimension reduction. In tandem, LSTM models, a type of recurrent neural network (RNN), were harnessed to capture long-term dependencies and effectively model sequential data, particularly time series. This fusion of CNN and LSTM networks aimed to significantly enhance the precision of short-term predictions. Notably, this study represents a pioneering effort, encompassing a comprehensive analysis of data from 24 countries worldwide. In a rigorous comparison with other deep learning methods, the proposed hybrid CNN-LSTM model demonstrated superior accuracy and adaptability, affirming its position as a formidable choice for predictive modeling in this context.

Furthermore, the hybrid GWO-WAO algorithm was employed to optimize the hyperparameters of the proposed model, including the number of filters, kernel size, pool size, and LSTM units. The WOA algorithm was compared against other optimization algorithms, namely GA, GWO, and FA, demonstrating its effectiveness in fine-tuning the model's hyperparameters. 
Additionally, a comparative analysis was conducted with other models including CNN, RCLSTM, BiLSTM, CNN-GRU, and ConvLSTM. Across various experiments, the proposed method consistently outperformed these models in terms of prediction accuracy.
 
Comprehensive comparison was conducted of the proposed model against various deep learning methods using rigorous statistical tests, including the Friedman and Nemenyi post-hoc test. The results unequivocally establish the hybrid CNN-LSTM as the top performer, outshining other well-regarded deep learning techniques. Notably, the exceptional accuracy achieved by our proposed method is attributed to a novel hybrid hyperparameter tuning approach and the distinctive framework of the CNN-LSTM hybrid model.\\
In future directions, further enhancement of the proposed method could be pursued by  exploring the impact of external data such as age, location, and sex could be investigated to improve the overall performance of the proposed method.

\printbibliography{}

\end{document}